\documentclass{bioinfo}
\copyrightyear{2010}
\pubyear{2010}

\newcommand{\dna}[1]{\hbox{\small \texttt{\uppercase{#1}}}}
\newcommand{\rank}{\mathop{\textit{rank}}\nolimits}
\newcommand{\select}{\mathop{\textit{select}}\nolimits}
\newcommand{\bigo}[1]{\ensuremath{O(#1)}}
\newcommand{\littleo}[1]{\ensuremath{o(#1)}}

\usepackage[hang,raggedright]{subfigure}
\usepackage{tikz}
\usetikzlibrary{snakes,arrows,shapes,matrix}

\begin{document}
\firstpage{1}

\title[Succinct Data Structures for Assembling Large Genomes]{Succinct Data Structures for Assembling Large Genomes}
\author[Conway \textit{et~al}]{Thomas C Conway$^{1,*}$ and Andrew J Bromage$^{1,}$\footnote{to whom correspondence should be addressed}}
\address{$^{1}$NICTA, Victoria Research Laboratory,\\
Department of Computer Science and Engineering, University of Melbourne,\\
Australia}

\history{Received on XXXXX; revised on XXXXX; accepted on XXXXX}

\editor{Associate Editor: XXXXXXX}

\maketitle

\begin{abstract}

\section{Motivation:}
Second generation sequencing technology makes it feasible for
many researches to obtain enough sequence reads to attempt
the \emph{de~novo} assembly of higher eukaryotes (including mammals).
\emph{De~novo} assembly not only provides a tool for understanding
wide scale biological variation, but within human bio-medicine,
it offers a direct way of observing both large scale structural
variation and fine scale sequence variation.
Unfortunately,
improvements in the computational feasibility for \emph{de~novo} assembly
have not matched the improvements in the gathering of sequence data.
This is for two reasons: the inherent computational complexity of the
problem, and the in-practice memory requirements of tools.

\section{Results:}
In this paper we use entropy compressed or \textit{succinct} data structures
to create a practical representation of the de~Bruijn assembly graph,
which requires at least a factor of 10 less storage
than the kinds of structures used by deployed methods.
In particular we show that when stored succinctly,
the de~Bruijn assembly graph for homo sapiens
requires only 23~gigabytes of storage.
Moreover, because our representation is entropy compressed,
in the presence of sequencing errors it has better scaling behaviour 
asymptotically than conventional approaches.

\section{Availability:}
Binaries of programs for constructing and traversing the de~Bruijn assembly graph
are available from\\
\href{http://www.genomics.csse.unimelb.edu.au/succinctAssembly}{http://www.genomics.csse.unimelb.edu.au/succinctAssembly}.

\section{Contact:} \href{tom.conway@nicta.com.au}{tom.conway@nicta.com.au}
\end{abstract}

\section{Introduction}

A central problem in sequence bioinformatics is that of assembling
genomes from a collection of overlapping short fragments thereof. These
fragments are usually the result of sequencing -- the determination by
an instrument of a sampling of subsequences present in a sample of DNA.
The number,
length and accuracy of these sequences, varies significantly between the
specific technologies, as does the degree of deviation from uniform sampling,
and all these are constantly changing as new technologies
are developed and refined. Nonetheless, it is typically the case
that we have anywhere from hundreds of thousands of sequences several
hundred bases in length to hundreds of millions of sequences a few tens
of bases in length with error rates between 0.1\% and 10\%, depending
on the technology.

The two main techniques used for reconstructing the underlying sequence
from the short fragments are based on overlap-layout-consensus models, and
de~Bruijn graph models. The former was principally used with older
sequencing technologies which tend to yield fewer longer reads, and
the latter has become increasingly popular with the so-called next
generation sequencing technologies which yield many more shorter
sequence fragments.
Irrespective of the technique, \cite{Med07} shows that the problem of
sequence assembly is computationally hard,
and as the correct solution is not rigorously defined,
all practical assembly techniques are necessarily
heuristic in nature. It is not our purpose here to discuss the various
assembly techniques -- we restrict our attention to certain
aspects of de~Bruijn graph assembly -- we refer the reader to
\cite{Miller10} for a fairly comprehensive review of assemblers and
assembly techniques.

Space consumption is a pressing practical problem for assembly with
de~Bruijn graph based algorithms (we have observed \textit{velvet}
using 20~GB to assemble staphylococcus aureus -- a 2.8~Mbp genome),
and we present a representation for the de~Bruijn assembly graph
that is extremely compact.
The representations we present use entropy compressed or \textit{succinct}
data structures.
These are representations, typically of sets or sequences of integers
that use an amount of space bounded closely by the theoretical minimum
suggested by the zero-order entropy of the set or sequence.
These representations combine their space efficiency with efficient access.
In some cases query operations can be performed in constant time,
and in most cases they are at worst logarithmic.

Succinct data structures are a basic building block;
\cite{Jacobson89} shows
more complex discrete data structures such as
trees and graphs can be built using them.
Some of the tasks for which they have been used include Web
graphs (\cite{ClaudeNavarro07}), XPath indexing (\cite{XPathSearch09}),
partial sums (\cite{Hon03}) and short read alignment (\cite{KiSu09}).

\subsection{de~Bruijn Graphs \& Assembly}

Let $\Sigma$ be an alphabet, and $|\Sigma|$ be the number of symbols
in that alphabet.
In the case of genome assembly,
the alphabet~$\Sigma$ is~$\{ \dna{a}, \dna{c}, \dna{g}, \dna{t} \}$.
The length of a string~$s$ of symbols drawn from~$\Sigma$
is written~$|s|$.
The notation $s[i,j)$ is used for the substring of $s$ starting at position
$i$ (counting from 0) to, but not including $j$.

The directed de~Bruijn graph of degree $k$ is defined as
\begin{eqnarray*}
    G_{*} &=& \left\langle V_{*}, E_{*} \right\rangle \\
    V_{*} &=& \left\{ s : s \in \Sigma^k \right\} \\
    E_{*} &=& \left\{ \left\langle n_f, n_t \right\rangle : n_f, n_t \in V_{*};
                n_f[1,k) = n_t[0,k-1) \right\}
\end{eqnarray*}

That is, the nodes of the de~Bruijn graph $V_{*}$ correspond to all the $k$ length strings
over $\Sigma$ and an edge exists between each pair of nodes for which the
last $k-1$ symbols of the first are the same as the first $k-1$ of the second.

The $k$ length string labelling a node is usually referred to as a $k$-gram
in the computer science literature and a $k$-mer in the bioinformatics literature.
The labels of the edges, as noted in \cite{Good46}, are $k+1$-mers.
For clarity, we use $\rho = k + 1$, and refer to edges as $\rho$-mers.

We note that amongst the special properties of the de~Bruijn
graph is the fact that a given node can have at most $|\Sigma|$ successor nodes:
formed by taking the last $k$ bases of the node and extending them with
each of the symbols in the alphabet.
That is, we can define the successors of a node $n$:

\begin{eqnarray}
    \textit{succ}_{*}(n) &=& \left\{ n[1,k) \cdot b : b \in \Sigma \right\}\label{eqn:succ} \\
    \textit{pred}_{*}(n) &=& \left\{ b \cdot n[0,k-1) : b \in \Sigma \right\}\label{eqn:pred}
\end{eqnarray}

To use the de~Bruijn graph for assembly, we can build a subset of the
graph by finding the nodes and edges that are supported by the information
in the sequence reads.
The edges are also annotated with a count of the number of times that
a $\rho$-mer is observed in the sequence data.
The counts are used for two purposes.
The first is to distinguish edges that arise from sequencing errors
(which will have very low counts) from those that arise from the underlying
genome (which will have higher counts).
The second is to estimate the number of copies of that edge in the underlying
genome.

Given a set of reads $S$, we can define a \textit{de~Bruijn assembly graph},
defining the nodes $V_S$ in terms of the
edges $E_S$ rather than the other way round, as we did above.
To define the nodes, we create two (overlapping) sets: the set of
nodes $F_S$ \textit{from} which an edge proceeds,
and the set of nodes $T_S$ \textit{to} which an edge proceeds.

\begin{eqnarray}
    E_{S} &=& \left\{ s_i[j,j + \rho) : 0 \le j < |s_i| - k; \forall s_i \in S \right\} \\
    F_{S} &=& \left\{ e[1,\rho+1) : e \in E_{S} \right\} \nonumber \\
    T_{S} &=& \left\{ e[0,\rho) : e \in E_{S} \right\} \nonumber \\
    V_{S} &=& F_{S} \cup T_{S} \\
    G_{S} &=& \left\langle V_{S}, E_{S} \right\rangle
\end{eqnarray}

From the DNA alphabet and equation~\ref{eqn:succ},
a given node in the assembly graph can have at most~4 successor nodes,
and by equation~\ref{eqn:pred},
a given node can also have at most~4 predecessor nodes.

\subsubsection{Reverse Complements}

An important distinction between ideal strings and the DNA sequences that
are used in genome assembly is that the latter can be read in two
directions: forwards, and in the reverse direction with the individual
DNA letters exchanged with their Watson-Crick complements
($\dna{a} \leftrightarrow \dna{t}$ and~$\dna{c} \leftrightarrow \dna{g}$).
In most sequencing scenarios, fragments of DNA are randomly sequenced in
either direction, something that must be taken into account during assembly.
First, sequence reads are processed twice -- once reading them forwards,
and then reading them in the reverse complement direction.
Then, in most cases, nodes corresponding to reverse complement
sequences are merged,
and the edges are made bi-directed to match up the sequences correctly
(see, for example \cite{Med07}).
For our current discussion, we will not combine them, but will store
them separately.
This makes the graph symmetric -- a forward traversal corresponds to
a backwards traversal on the reverse complement path, and vise versa.

\begin{figure*}
\caption{A de~Bruijn assembly graph and its representation.}\label{fig:graph}

\begin{tabular}{ll}
\begin{minipage}{0.7\textwidth}
\centering
\subfigure[Source sequence]{
\begin{tabular}{ll}\toprule
Sequence: & \dna{gctttcgacgtttca} \\
Reverse complement: & \dna{tgaaacgtcgaaagc} \\\botrule
\end{tabular}
}

\subfigure[The corresponding assembly graph. The edges are labelled with counts.
           The path marked with bold arrows is \dna{ttcgac}.]{
\begin{tikzpicture}[>=latex',scale=0.6]
  \node (agc) at (0,3) [draw,rectangle] {\dna{agc}};
  \node (aag) at (2,3) [draw,rectangle] {\dna{aag}};
  \node (aaa) at (4,3) [draw,rectangle] {\dna{aaa}};
  \node (aac) at (5,5) [draw,rectangle] {\dna{aac}};
  \node (acg) at (7,5) [draw,rectangle] {\dna{acg}};
  \node (cgt) at (9,5) [draw,rectangle] {\dna{cgt}};
  \node (gtt) at (11,5) [draw,rectangle] {\dna{gtt}};
  \node (gaa) at (5,1) [draw,rectangle] {\dna{gaa}};
  \node (cga) at (7,1) [draw,rectangle] {\dna{cga}};
  \node (tcg) at (9,1) [draw,rectangle] {\dna{tcg}};
  \node (ttc) at (11,1) [draw,rectangle] {\dna{ttc}};
  \node (tga) at (3,0) [draw,rectangle] {\dna{tga}};
  \node (tca) at (13,0) [draw,rectangle] {\dna{tca}};
  \node (gac) at (7,3) [draw,rectangle] {\dna{gac}};
  \node (gtc) at (9,3) [draw,rectangle] {\dna{gtc}};
  \node (ttt) at (12,3) [draw,rectangle] {\dna{ttt}};
  \node (ctt) at (14,3) [draw,rectangle] {\dna{ctt}};
  \node (gct) at (16,3) [draw,rectangle] {\dna{gct}};
  \draw [->] (acg) edge[] node[above] {2} (cgt);
  \draw [->] (aac) edge[] node[above] {1} (acg);
  \draw [->] (gaa) edge[] node[right] {2} (aaa);
  \draw [->] (ttt) edge[] node[left] {2} (ttc);
  \draw [->] (aaa) edge[] node[left] {1} (aac);
  \draw [->] (ctt) edge[] node[above] {1} (ttt);
  \draw [->] (ttc) edge[] node[above] {1} (tca);
  \draw [->] (cgt) edge[] node[above] {1} (gtt);
  \draw [->] (cgt) edge[] node[left] {1} (gtc);
  \draw [->,very thick] (ttc) edge[] node[above] {1} (tcg);
  \draw [->] (cga) edge[] node[above] {1} (gaa);
  \draw [->] (tga) edge[] node[above] {1} (gaa);
  \draw [->] (gac) edge[] node[left] {1} (acg);
  \draw [->] (aag) edge[] node[above] {1} (agc);
  \draw [->,very thick] (cga) edge[] node[left] {1} (gac);
  \draw [->] (gtc) edge[] node[left] {1} (tcg);
  \draw [->] (aaa) edge[] node[above] {1} (aag);
  \draw [->] (gtt) edge[] node[right] {1} (ttt);
  \draw [->,very thick] (tcg) edge[] node[above] {2} (cga);
  \draw [->] (gct) edge[] node[above] {1} (ctt);
\end{tikzpicture}
}
\end{minipage} &
\begin{minipage}{0.2\textwidth}
\subfigure[Extracted $\rho$-mers with counts.]{
\begin{tabular}{ll|ll}\toprule
$\rho$-mer & count & $\rho$-mer & count \\ \midrule
\dna{aaac} & 1 & \dna{gaaa} & 2 \\
\dna{aaag} & 1 & \dna{gacg} & 1 \\
\dna{aacg} & 1 & \dna{gctt} & 1 \\
\dna{aagc} & 1 & \dna{gtcg} & 1 \\
\dna{acgt} & 2 & \dna{gttt} & 1 \\
\dna{cgaa} & 1 & \dna{tcga} & 2 \\
\dna{cgac} & 1 & \dna{tgaa} & 1 \\
\dna{cgtc} & 1 & \dna{ttca} & 1 \\
\dna{cgtt} & 1 & \dna{ttcg} & 1 \\
\dna{cttt} & 1 & \dna{tttc} & 2 \\\botrule
\end{tabular}
}
\end{minipage}
\end{tabular}

\subfigure[The sparse bitmap representation.
           The gray boxes exemplify groups of edges that proceed from a single node.
           The arrows show the sequence \dna{ttcgac}.]{
\begin{tikzpicture}
 \matrix(m)[matrix, row sep=0.1cm, column sep=-0.05cm]{
\node(aaaa) {0}; \pgfmatrixnextcell
\node(aaac) {1}; \pgfmatrixnextcell
\node(aaag) {0}; \pgfmatrixnextcell
\node(aaat) {0}; \pgfmatrixnextcell
\node(aaca) {0}; \pgfmatrixnextcell
\node(aacc) {0}; \pgfmatrixnextcell
\node(aacg) {1}; \pgfmatrixnextcell
\node(aact) {0}; \pgfmatrixnextcell
\node {$\ldots$}; \pgfmatrixnextcell
\node(acgt) {1}; \pgfmatrixnextcell
\node {$\ldots$}; \pgfmatrixnextcell
\node(cgaa) {1}; \pgfmatrixnextcell
\node(cgac) {1}; \pgfmatrixnextcell
\node(cgag) {0}; \pgfmatrixnextcell
\node(cgat) {0}; \pgfmatrixnextcell
\node {$\ldots$}; \pgfmatrixnextcell
\node(cgtc) {1}; \pgfmatrixnextcell
\node(cgtg) {0}; \pgfmatrixnextcell
\node(cgtt) {1}; \pgfmatrixnextcell
\node {$\ldots$}; \pgfmatrixnextcell
\node(cttt) {1}; \pgfmatrixnextcell
\node(gaaa) {1}; \pgfmatrixnextcell
\node {$\ldots$}; \pgfmatrixnextcell
\node(gaca) {0}; \pgfmatrixnextcell
\node(gacc) {0}; \pgfmatrixnextcell
\node(gacg) {1}; \pgfmatrixnextcell
\node(gact) {0}; \pgfmatrixnextcell
\node {$\ldots$}; \pgfmatrixnextcell
\node(gctt) {1}; \pgfmatrixnextcell
\node {$\ldots$}; \pgfmatrixnextcell
\node(gttt) {1}; \pgfmatrixnextcell
\node {$\ldots$}; \pgfmatrixnextcell
\node(tcga) {1}; \pgfmatrixnextcell
\node(tcgc) {0}; \pgfmatrixnextcell
\node(tcgg) {0}; \pgfmatrixnextcell
\node(tcgt) {0}; \pgfmatrixnextcell
\node(tgaa) {1}; \pgfmatrixnextcell
\node {$\ldots$}; \pgfmatrixnextcell
\node(ttca) {1}; \pgfmatrixnextcell
\node(ttcc) {0}; \pgfmatrixnextcell
\node(ttcg) {1}; \pgfmatrixnextcell
\node(ttct) {0}; \pgfmatrixnextcell
\node {$\ldots$}; \pgfmatrixnextcell
\node(ttta) {0}; \pgfmatrixnextcell
\node(tttc) {1}; \pgfmatrixnextcell
\node(tttg) {0}; \pgfmatrixnextcell
\node(tttt) {0}; \\
};
\node [below,anchor=mid east,rotate=90] at (aaaa.south) {\dna{aaaa}};
\node [below,anchor=mid east,rotate=90] at (aaac.south) {\dna{aaac}};
\node [below,anchor=mid east,rotate=90] at (aaag.south) {\dna{aaag}};
\node [below,anchor=mid east,rotate=90] at (aaat.south) {\dna{aaat}};
\node [below,anchor=mid east,rotate=90] at (aaca.south) {\dna{aaca}};
\node [below,anchor=mid east,rotate=90] at (aacc.south) {\dna{aacc}};
\node [below,anchor=mid east,rotate=90] at (aacg.south) {\dna{aacg}};
\node [below,anchor=mid east,rotate=90] at (aact.south) {\dna{aact}};
\node [below,anchor=mid east,rotate=90] at (acgt.south) {\dna{acgt}};
\node [below,anchor=mid east,rotate=90] at (cgaa.south) {\dna{cgaa}};
\node [below,anchor=mid east,rotate=90] at (cgac.south) {\dna{cgac}};
\node [below,anchor=mid east,rotate=90] at (cgag.south) {\dna{cgag}};
\node [below,anchor=mid east,rotate=90] at (cgat.south) {\dna{cgat}};
\node [below,anchor=mid east,rotate=90] at (cgtc.south) {\dna{cgtc}};
\node [below,anchor=mid east,rotate=90] at (cgtg.south) {\dna{cgtg}};
\node [below,anchor=mid east,rotate=90] at (cgtt.south) {\dna{cgtt}};
\node [below,anchor=mid east,rotate=90] at (cttt.south) {\dna{cttt}};
\node [below,anchor=mid east,rotate=90] at (gaaa.south) {\dna{gaaa}};
\node [below,anchor=mid east,rotate=90] at (gaca.south) {\dna{gaca}};
\node [below,anchor=mid east,rotate=90] at (gacc.south) {\dna{gacc}};
\node [below,anchor=mid east,rotate=90] at (gacg.south) {\dna{gacg}};
\node [below,anchor=mid east,rotate=90] at (gact.south) {\dna{gact}};
\node [below,anchor=mid east,rotate=90] at (gctt.south) {\dna{gctt}};
\node [below,anchor=mid east,rotate=90] at (gttt.south) {\dna{gttt}};
\node [below,anchor=mid east,rotate=90] at (tcga.south) {\dna{tcga}};
\node [below,anchor=mid east,rotate=90] at (tcgc.south) {\dna{tcgc}};
\node [below,anchor=mid east,rotate=90] at (tcgg.south) {\dna{tcgg}};
\node [below,anchor=mid east,rotate=90] at (tcgt.south) {\dna{tcgt}};
\node [below,anchor=mid east,rotate=90] at (tgaa.south) {\dna{tgaa}};
\node [below,anchor=mid east,rotate=90] at (ttca.south) {\dna{ttca}};
\node [below,anchor=mid east,rotate=90] at (ttcc.south) {\dna{ttcc}};
\node [below,anchor=mid east,rotate=90] at (ttcg.south) {\dna{ttcg}};
\node [below,anchor=mid east,rotate=90] at (ttct.south) {\dna{ttct}};
\node [below,anchor=mid east,rotate=90] at (ttta.south) {\dna{ttta}};
\node [below,anchor=mid east,rotate=90] at (tttc.south) {\dna{tttc}};
\node [below,anchor=mid east,rotate=90] at (tttg.south) {\dna{tttg}};
\node [below,anchor=mid east,rotate=90] at (tttt.south) {\dna{tttt}};
\draw[overlay,very thick,opacity=0.5] (ttca.north west) rectangle (ttct.south east);
\draw[overlay,very thick,opacity=0.5] (tcga.north west) rectangle (tcgt.south east);
\draw[overlay,very thick,opacity=0.5] (cgaa.north west) rectangle (cgat.south east);
\draw[overlay,very thick,opacity=0.5] (gaca.north west) rectangle (gact.south east);
\draw[overlay,->,very thick,opacity=0.5] (ttcg.north) to[bend right] (tcgt.north east);
\draw[overlay,->,very thick,opacity=0.5] (tcga.north) to[bend right=20] (cgat.north east);
\draw[overlay,->,very thick,opacity=0.5] (cgac.north) to[bend left] (gaca.north west);
\end{tikzpicture}
}

\subfigure[Dense array of counts.]{
\begin{tikzpicture}
 \matrix(m)[matrix, row sep=0.1cm, column sep=0.0cm]
  {
\node {1}; \pgfmatrixnextcell
\node {1}; \pgfmatrixnextcell
\node {1}; \pgfmatrixnextcell
\node {1}; \pgfmatrixnextcell
\node {2}; \pgfmatrixnextcell
\node {1}; \pgfmatrixnextcell
\node {1}; \pgfmatrixnextcell
\node {1}; \pgfmatrixnextcell
\node {1}; \pgfmatrixnextcell
\node {1}; \pgfmatrixnextcell
\node {2}; \pgfmatrixnextcell
\node {1}; \pgfmatrixnextcell
\node {1}; \pgfmatrixnextcell
\node {1}; \pgfmatrixnextcell
\node {1}; \pgfmatrixnextcell
\node {2}; \pgfmatrixnextcell
\node {1}; \pgfmatrixnextcell
\node {1}; \pgfmatrixnextcell
\node {1}; \pgfmatrixnextcell
\node {2}; \\
  };
\end{tikzpicture}
}
\end{figure*}
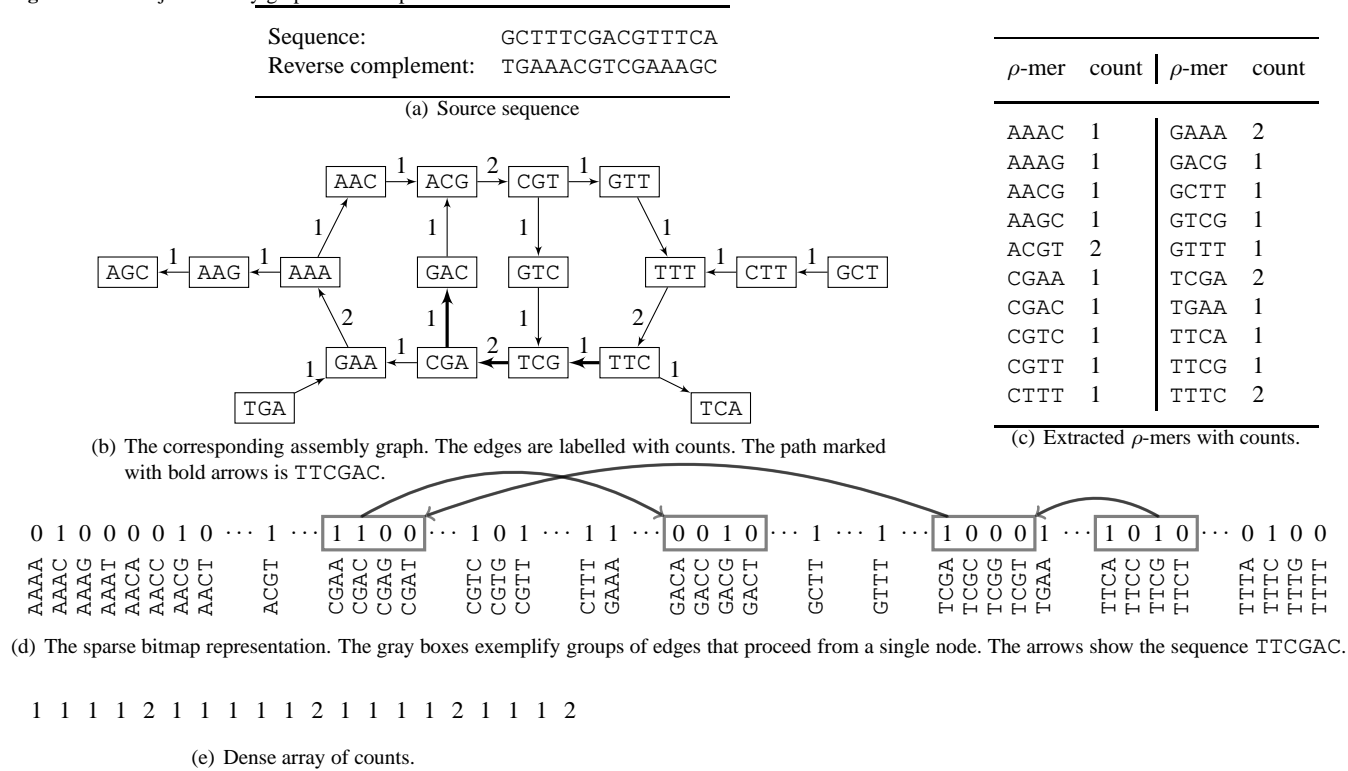

Figure~\ref{fig:graph} shows a de~Bruijn assembly graph for a short string.

\subsubsection{From de~Bruijn Assembly Graphs To Genomes}

The de~Bruijn graph is both Eulerian and Hamiltonian, a property that
\cite{IduryWaterman95} showed was useful for genome assembly.
In principle, at least, the assembled sequence corresponds to an Eulerian tour
of the de~Bruijn assembly graph.
The details of how this may be done in practice are beyond the
scope of our current discussion, but the approaches include those
described in \cite{Pevzner01,Velvet,Jackson2009,Simpson2009}.
Our current discussion is focussed on how we might represent
the de~Bruijn assembly graph in a practical program for performing
large genome assembly.

A simple approach to representing the de~Bruijn assembly graph is to
represent the nodes as ordinary records (i.e. using a \texttt{struct}
in~C or C++), and the edges as pointers between them.
If we assume a node contains the $k$ length substring (or $k$-mer)
represented as a 64 bit integer (assuming $k \le 32$),
32~bit edge counts and pointers to four possible successor nodes,
and there are no memory allocator overheads,
then the graph will require 56 bytes per node.
In the drosophila melanogaster genome, with $k=25$,
there are about 245 million nodes (including reverse complements),
so we would expect the graph to take nearly 13~GB.
For the human genome with $k=25$ there are about 4.8 billion
nodes (again, including reverse complements), so the graph
would require over~250~GB.
These data structures are large, but more is needed, because there
is no way in what is described to locate a given node, so for instance
a simple hash table (generously assuming a load factor of~1) might
require an extra~16 bytes (hash value + pointer) per node or over~70~GB
for the human genome.
These figures are extremely conservative,
since they ignore the effect of sequencing errors.

We can get an estimate of the proportion of edges in the
graph that are due to errors with a simple analysis.
Most sequencing errors give
rise to unique $k$-mers, and hence many edges that occur only once.
Ignoring insertion and deletion errors,
for a given $k$ (or $\rho$), a single error leads to $\rho$
spurious edges, which, if we assume a random distribution of errors,
are overwhelmingly likely to be unique.
Thus, the number of spurious edges is proportional
to the volume of sequence data, whereas the number of true edges
is proportional to the genome size, and will converge on that
number as the volume of sequence data increases.
For example, consider the case of an organism with a 1~Mbp genome,
which we sequence with sequence reads 100bp in length.
If we assume that on average a read contains~1 error, then with
$\rho=26$, we will typically have~74 true edges and~26 spurious
edges.
Assuming the reads are uniformly distributed, once the number of reads
exceeds about~14,000, almost all the~1~million true edges will be present,
and there will be about~364,000 spurious edges. Beyond this, as the
number of reads increases, the number of true edges will remain the
same, but the number of spurious edges will continue to increase linearly.
By the time the \textit{coverage} (the expected count on all the true
edges) reaches~40 (a typical coverage for genome assembly),
we would expect to see about~14~million spurious edges.
That is, the spurious edges would outnumber the true edges by a factor of~14.

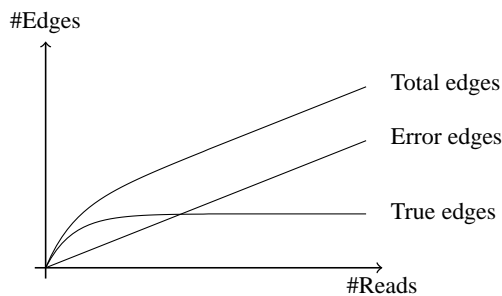
\begin{figure}
\begin{tikzpicture}[x=5cm/7,y=5cm/7]
\draw[->, semithick] (-0.2,0) -- (6.2,0) node[below] {\#Reads};
\draw[->, semithick] (0,-0.2) -- (0,4.2) node[above] {\#Edges};
\pgfplothandlerlineto
\pgfplotfunction{\x}{0,0.1,...,6.0}{\pgfpointxy{\x}{(1 - (\x<3)*exp(-2*\x))}}
\pgfusepath{stroke}\node at (6.2,1.0) [right] {True edges};

\pgfplothandlerlineto
\pgfplotfunction{\x}{0,0.1,...,6.0}{\pgfpointxy{\x}{0.4*\x}}
\pgfusepath{stroke}\node at (6.2,2.4) [right] {Error edges};

\pgfplothandlerlineto
\pgfplotfunction{\x}{0,0.1,...,6.0}{\pgfpointxy{\x}{0.4*\x + (1 - (\x<3)*exp(-2*\x))}}
\pgfusepath{stroke}\node at (6.2,3.4) [right] {Total edges};

\end{tikzpicture}
\caption{A sketch showing the relationship between the number of
sequence reads and the number of edges in the graph.
Because the underlying genome is fixed in size,
as the number of sequence reads increases the number of edges in the
graph due to the underlying genome will plateau when every part of
the genome is covered. Conversely, since errors tend to be random
and more-or-less unique, their number scales linearly with the number
of sequence reads. Once enough sequence reads are present to have
enough coverage to clearly distinguish true edges (which come from the
underlying genome), they will usually be outnumbered by spurious edges
(which arise from errors) by a substantial factor.
}
\label{fig:spuriousedges}
\end{figure}

Figure~\ref{fig:spuriousedges} illustrates this problem.

Much of this space is devoted to storing pointers,
so the question naturally arises: are these pointers necessary,
or can they be avoided?
Existing assemblers such as \textit{velvet} (\cite{Velvet})
and \textit{ABySS} (\cite{Simpson2009})
combine nodes corresponding to forward and reverse complements,
and merge nodes on unbranched paths,
and although these techniques significantly reduce the amount of
memory required, they none the less represent an \emph{ad hoc} approach
to the problem of reducing the memory required to represent the
de~Bruijn assembly graph.

\section{Approach}
\label{sec:approach}

Our approach to memory-efficient representation of an assembly graph
begins by reframing the question of whether the pointers
in a naive graph representation are necessary.
Rather we ask what information is necessary,
and what is redundant or ephemeral.
How many bits are required to represent the de~Bruijn assembly graph
from an information-theoretic point of view?

The de~Bruijn assembly graph is a subset of the de~Bruijn graph.
Of the $|\Sigma|^\rho$ edges in the de~Bruijn graph,
the assembly graph contains $|E_S|$.
The self-information of a set of edges that make up an assembly graph, and
hence the minimum number of bits required to encode the graph, is
\begin{eqnarray}
    \#\textit{bits} &=& \log {{4^\rho} \choose {\left|E_S\right|}}\label{eqn:bits}
\end{eqnarray}
(Note, that unless otherwise specified, all logarithms are base 2.)

For the de~Bruijn assembly graph with $k = 25$,
the human genome (build 37) yields 4,796,397,453
distinct edges, including reverse complements.
By the equation above, taking $S$ to be the genome itself:
\[
    \#\textit{bits} = \log {{4^{26}} \choose {4,796,397,453}} \approx 1.2~\textrm{GB}
\]

We do not need to store the nodes explicitly, since they are
readily inferred from the edges:
\begin{eqnarray*}
    \textit{from-node}(e) &=& e[0,\rho - 1) \\
    \textit{to-node}(e) &=& e[1,\rho)
\end{eqnarray*}

Equation~\ref{eqn:bits} gives a lower bound on the number of
bits required to represent the de~Bruijn assembly graph.
We would like to find a concrete representation that comes close
to that theoretical minimum while allowing efficient random access.
The notion that the assembly graph is a subset of the de~Bruijn
graph immediately suggests that we could create a bitmap with a
bit for each edge in the de~Bruijn graph, and set the bits for
the edges that occur in the assembly graph.
Such a scheme depends on being able to enumerate the $\rho$-mers
(i.e. the edges).
This is done trivially by numbering the bases (we use $\dna{a} = 0$,
$\dna{c} = 1$, $\dna{g} = 2$ and~$\dna{t} = 3$),
and interpreting the $\rho$ symbols as an integer with $2\rho$~bits.
Conceptually, then, we can create a bitmap with $4^\rho$~bits, and place
\textbf{1}s in the positions corresponding to the edges in the assembly
graph.

Given such a bitmap,
we can determine the successor set of a given node from the
definition of the de~Bruijn assembly graph,
by probing the positions corresponding to the 4 edges that could
proceed from the node. For a node corresponding to a $k$-mer $n$
the four positions in the bitmap are $4n$, $4n + 1$, $4n + 2$ and $4n + 3$.

There is a particular formalism, first proposed by \cite{Jacobson89} for querying sets of integers represented as bitmaps
which is useful in this setting. 
Given a bitmap $\mathbf{b}$ with the positions of the
set members set to \textbf{1} and the rest of the positions set to \textbf{0},
the formalism uses two query operators
\textit{rank} and \textit{select} with the following definitions\
\footnote{The literature contains several slightly different definitions
that arise from different conventions for subscripting arrays --
mathematical literature tends subscript from one; computer science literature
from zero. We use the latter.}:

\begin{eqnarray*}
\rank_{\mathbf{b}}(p) & = & \sum_{0 \le i < p} b_i \\
\select_{\mathbf{b}}(i) & = & \max \left\{ p < n | \rank_{\mathbf{b}}(p) \le i \right\}
\end{eqnarray*}

Intuitively, $\rank_{\mathbf{b}}(p)$ is the number of ones in the bitmap
$\mathbf{b}$ to the left of position $p$, and $\select_{\mathbf{b}}(i)$
is the position of the~$i$-th set bit, where the set bits are numbered
starting from zero.

These operations are inverses in that $\rank_{\mathbf{b}}(\select_{\mathbf{b}}(i)) = i$
for~$i\in \left\{ 0 \ldots \mu-1 \right\}$, and $\select_{\mathbf{b}}(\rank_{\mathbf{b}}(p)) = p$
for~$p \in \left\{ p : p \in \{ 0 \ldots \nu-1 \}; b_p = \mathbf{1} \right\}$.

Using the \textit{rank}/\textit{select} formalism, we can
compute the set of the ranks of the successor edges for a node $n$ efficiently
given a bitmap representing the set of edges:
\begin{eqnarray*}
    \textit{succ}_{S}(n) &=& \left\{ r \in \left[\rank_{E_S}(4n), \rank_{E_S}(4n + 4)\right) \right\}
\end{eqnarray*}
This forms the basis of a method for efficient traversal of a de~Bruijn
assembly graph represented as a set of integers (i.e., a bitmap).

Next we consider how the edge counts should be represented.
For this we draw on the \textit{rank}/\textit{select} formalism again,
and note that while the edges are sparse (a point that we will come
back to shortly), the \textit{ranks} of the edges are dense,
filling the range $[0, \left|E_S\right|)$.
Therefore we can store the edge counts in a vector of 32~bit integers.

\section{Methods}

The preceding discussion presented a technique for representing a de
Bruijn assembly graph as a bitmap using $4^\rho$~bits.
For a typical value of $\rho = 26$ (i.e. $k = 25$), the bitmap would
require 512TB. This is clearly infeasible (and larger $k$ would be worse),
but the bitmap is incredibly sparse. Of the~$4.5\times10^{15}$
bits, for the human genome, only~$4.7\times10^{9}$ are~\textbf{1}.
That is, the fraction of the bits that are set is~$10^{-8}$,
so a sparse representation should be used.
In fact, Equation~\ref{eqn:bits} gives a precise
lower bound on the number of bits that a sparse representation requires,
and there has been a large amount of research in the last two decades
on the efficient representation of data structures that are close
to the theoretical limit.

\begin{table*}
\caption{Summary of succinct data structures.\label{tbl:succinctds}}
\begin{tabular}{llll}\toprule
Method & Size (bits) & rank complexity & select complexity \\\midrule
\textbf{darray} & $\nu + \littleo{\nu}$ & $\bigo{1}$ & $\bigo{\log^4 \mu / \log \nu}$ \\
\textbf{sarray} & $\mu \log \frac{\nu}{\mu} + 1.92\mu + \littleo{\mu}$ & $\bigo{\log \frac{\nu}{\mu}} + \bigo{\log^4 \mu / \log \nu}$ & $\bigo{\log^4 \mu / \log \nu}$ \\
\textbf{rrrarray} & $\log {\nu \choose \mu} + \littleo{\mu} + \bigo{\log \log \nu}$ & 
	$\bigo{1}$ & $\bigo{1}$ \\\botrule
\end{tabular}
\end{table*}

Let $\mathcal{B}_{\nu,\mu}$ be the set of bitmaps with $\nu$~bits,
where exactly $\mu$~bits are set.  \cite{Jacobson89} defines a \emph{succinct
representation} as a way of mapping the elements
of $\mathcal{B}_{\nu,\mu}$ into a read-only memory such that
the amount of space used to represent a bitmap is close to
$(1 + \littleo{1}) \log \left|\mathcal{B}_{\nu,\mu}\right|$ bits.
A \emph{succinct data structure} is a succinct representation which
also supports desired query operations efficiently.
``Efficiently'' can mean either low asymptotic complexity,
or practical speed on real hardware.
In our case, the query operations that we wish to
support are \textit{rank} and \textit{select}.
Although \cite{Jacobson89} defines succinct data structures as read-only
objects, \cite{RRR01} and \cite{MakNav08}, amongst others,
show how \textit{insert} and \textit{delete} can be implemented without
sacrificing the succinct nature of the representation.
A summary (abstracted from \cite{OkSa06}) of the data structures that we use are shown in
Table~\ref{tbl:succinctds}.

The \textbf{darray} and \textbf{sarray} data structures~(\cite{OkSa06})
are optimised for the case when the bitmap is ``dense'' or ``sparse''
respectively.  If $\mu \approx \nu/2$, $\log {\nu \choose \mu} \approx \nu$,
so storing the uncompressed bitmap is optimal; in this case, the bitmap
is dense, and so \emph{rank} and \emph{select} can be implemented
with $\littleo{\nu}$ extra space to speed up those operations.
If $\mu/\nu$ is small, then the bitmap is sparse; in this case, the
bitmap can be compressed close to optimal space using Elias-Fano
coding (\cite{Elias74}), which is the basis for \textbf{sarray}.
The other main data structure that we use is \textbf{rrrarray}~(\cite{RRR07}),
which uses space very close to optimal over the entire range of values
of $\mu/\nu$, with moderate overhead in space usage.

To create a representation of the de~Bruijn assembly graph,
we extract the $\rho$-mers from the input sequences,
accumulating them in RAM and when RAM is ``full'', sorting them
(retaining duplicates) and writing the sorted run to disk.
Once all the $\rho$-mers have been accumulated into sorted runs,
the runs are then binary merged,
and the final list is processed, counting duplicate
$\rho$-mers to yield
a sequence of $\langle\textit{edge}, \textit{count}\rangle$ pairs
which are used to construct a sparse array (i.e. \textbf{sarray}),
and a vector of edge counts.
The merging phase uses $\log N$ passes, merging pairs of sorted runs.

Returning to the representation of the edge counts,
in Section~\ref{sec:approach}, we suggested storing the counts
in a vector of~32~bit integers indexed by edge rank.
This actually uses much more memory than necessary.
As previously noted, prior to error removal, a vast
majority of edges in the graph are spurious and will have a very low
edge count.
Most of the true edges have modest counts also: edges that are unique
in the underlying genome will have a count somewhere around the basic
coverage (e.g. 15--50).
For most edges 8~bits of storage is sufficient, and for most of the
remainder 16~bits is sufficient. Only a handful of edges, in practice,
need more than 16~bits.
Therefore using 32~bits for every edge is very wasteful.

There are many techniques for creating compressed representations
of vectors of integers (see~\cite{Moffat2002}), but in most
cases they do not provide efficient random access.
Succinct data structures implementing \textit{rank}/\textit{select}
yield an effective technique first introduced by \cite{Brisaboa09}.
We split each count into the three parts alluded to above: the least significant
8~bits, the ``middle'' 8~bits and the most significant 16~bits.
We store the least significant 8~bits in a dense vector of bytes~$L$.
Corresponding to it, we store a succinct bitmap~$B_L$
with a \textbf{1} marking
those entries for which the middle 8 bits, or the most significant 16 bits
are nonzero.
In a dense vector of bytes~$M$ (indexed by rank in~$B_L$) we store the
middle 8~bits of those entries for which a~\textbf{1} exists in~$B_L$.
Corresponding to~$M$, we store a sparse bitmap $B_M$ with a \textbf{1}
marking those entries for which the most significant 16 bits are nonzero.
Finally, we have a dense vector if 16~bit words~$H$
(indexed by rank in~$B_M$) with the most significant bits of those entries
marked in~$B_M$.
The bitmaps~$B_L$ and~$B_M$ are represented using \textbf{rrrarray}.


\section{Results}

\begin{table}
\caption{Genomes used for testing with the number of distinct edges
(excluding reverse complements, but including duplicates) as a
measure of the genome size,
the size of the assembly graph (including the edge counts) in bytes,
and the time, in seconds, taken to build the graph.\label{tbl:org}}
\begin{tabular}{lllr}\toprule
Organism &  \# Edges & Size & Time\\\midrule
mycobacterium leprae        & $3.2\times10^{6}$ & $3.3\times10^7$ & 5\\   
thalassiosira pseudonana    & $3.1\times10^{7}$ & $3.2\times10^8$ & 50\\   
caenorbahditis elegans      & $1.0\times10^{8}$ & $9.8\times10^8$ & 154\\   
arabidopsis thaliana        & $1.2\times10^{8}$ & $1.2\times10^9$ & 187\\   
drosophila melanogaster     & $1.6\times10^{8}$ & $1.3\times10^9$ & 317\\   
oryza sativa                & $3.7\times10^{8}$ & $3.0\times10^9$ & 428\\   
danio rerio                 & $1.5\times10^{9}$ & $1.1\times10^{10}$ & 5,448\\   
mus musculus                & $2.5\times10^{9}$ & $2.2\times10^{10}$ & 18,546\\   
homo sapiens                & $2.9\times10^{9}$ & $2.5\times10^{10}$ & 14,235\\\botrule   
\end{tabular}
\end{table}

We have created a program that uses straightforward implementations of
the succinct data structures we have described
to build de~Bruijn assembly graphs for the
genomes of the organisms listed in Table~\ref{tbl:org}.
All the reference genomes were obtained from the NCBI archive.
The number of edges (which include duplicates, but exclude reverse complements) are
marginally different to the genome sizes reported in the literature
(which themselves vary) because the edges do not include undetermined
bases represented as~\dna{n}s in the genomes,
and the size of the genome builds do not correspond exactly
to the estimated genome size.

In Table~\ref{fig:org} we report the size of the graph and multiplicities
data for the de~Bruijn assembly graph constructed over the reference genomes.
We also report the graph construction time on a server with
8 cores running at about 2~GHz, with 32~GB RAM, of which the graph construction
process used about 2~GB.
We consistently find that our code results in graphs requiring about 5.2 bytes per edge,
including the storage for the edge multiplicities which is less than
the 8 bytes for storing single pointer on a 64-bit architecture.
There is a greater degree of variability in the running time than there is
in the sizes of the assembly graphs, with the two largest genomes (mouse
and human) being the slowest (when weighted by genome size).
This is partly due to the $\log N$ on-disk passes required by the binary merge
used for the graph construction, and also because the disks were shared
on a cluster, and the longer runs will have suffered some degree of
interference.

It is difficult to compare fairly and directly to existing tools,
but to give some comparison, we tried running \textit{velvet}
(also with $k=25$).
Using synthesized error free reads,
at 30 times coverage, our 32~GB server was only able to assemble
the smallest of these genomes (mycobacterium leprae which is about 3.2~Mbp)
with an observed peak memory usage of about 325~MB.
The next smallest (thalassiosira pseudonana) failed when
the process (\texttt{velvetg}) exhausted main memory.
The graph and edge counts using our representation required 32~MB.

To test the speed of the random access in the graph,
we used a program that performs depth first traversal to find
the set of paths in the graph that do not contain branches.
We ran it on the graph for the thalassiosira organism with $k = 25$,
which contains 60,312,974 edges, and it took 202 seconds.
Each node is visited approximately once, and at each node the
code does 4 \textit{rank} operations to determine the number of
incoming and outgoing edges. Thus, the program is performing approximately
1.2 million \textit{rank} operations per second.
It is rather difficult to estimate how this would compare to a pointer
based implementation, but we would expect a pointer based implementation
to be up to an order of magnitude faster.
We note, however, that our implementation has not been highly tuned,
and in any case, the compactness of our representation makes the
thalassiosira genome fit in memory (easily), requiring 302~MB for
the graph and about 4~MB for the remaining structures required for the
graph traversal.

\section{Discussion}

The analysis, presented in Section~\ref{sec:approach},
suggested that for the human genome,
we would require a minimum of~1.2~GB, but
in our representation, we use~20~GB.
We expect the indexes that support the rank and select operations
to take some space, but the difference is more than an order of
magnitude.
The explanation lies in an important detail of the implementation
of the sparse array from \cite{OkSa06}.
The minimum space consumption calculation has two parameters:
$\nu$, the number of positions in the bitmap;
and $\mu$ the number of positions set to \textbf{1}.
In our computation we took $\nu = 4^\rho$, but the implementation,
which is (necessarily) built around machine word sizes takes
$\nu = 2^{64}$. If we recompute the minimum number of bits required
under that assumption, we have:
\[
    \#\textit{bits} = \log {{2^{64}} \choose {4,796,397,453}} \approx 19~\textrm{GB}
\]
Further research is necessary to
build an implementation of the sparse array that allows us
to set $\mu$ at values closer to $4^\rho$ when $\rho$ is less than 32.
It may be possible at values of $\nu = 2^{i + j}$ where $i$ and $j$
are machine words sizes (e.g. $i = 32$ and $j = 16$ would allow $\rho = 24$).

The techniques we have presented are by no means the only way to
reduce the memory requirements of de~Bruijn graph assembly.
Another approach is to use a hash table that maps from $k$-mer to a
record containing the counts on the 4 possible successor edges.
Since the 4 successor $k$-mers are trivially derived from the current
$k$-mer, we can store just this map. Moreover, this technique could use
a variable length coding technique like the one we use to store a single
byte for each of the counts in most cases.
For such a structure to be useful, we would need an open addressing
hash table, to avoid an indirection layer of pointers (as would be required
for a separate chaining hashing scheme).
Further, for a hashing scheme to be competitive, 
we would need to get the load factor close to 1.0.
Exactly such a hashing scheme, called \textit{cuckoo hashing}
has been proposed by \cite{Pagh01},
with several refinements, including those proposed by \cite{Ross07},
and \cite{Fotakis03}. The latter in particular shows a variant
that allows the load factor of the hash table to approach 1.0 while
retaining efficient access.
To make a rough comparison to our approach, let us consider the human
genome with 4.7 billion edges. The hashing scheme we have outlined
uses 8 bytes for the $k$-mer, and 1 byte for each of the counts. Since
only a tiny minority of edges will require more than 1 byte,
this will give us a close approximation to the space usage.
Thus, each edge requires 12 bytes. Assuming a load factor of 90\%,
which is likely to be close to the upper limit in practice,
this representation would require about 60~GB, which is a significant
improvement on the pointer based implementation, but is not nearly
as efficient as the representation we have proposed.

An important property of our representation, compared to either
of those outlined above, is that ours is \textit{succinct}.
That is, in a formal sense, ours uses within a small constant
factor, the minimum possible space.

We have presented a practical and efficient representation of the
de~Bruijn assembly graph, but of course there is much more to
doing \emph{de~novo} assembly with de~Bruijn graph methods.
Although we have sketched the space issues caused by sequencing errors,
we have not addressed the detection and correction of errors.
Also, a combinatoric number of Eulerian paths exist in the de~Bruijn
assembly graph, among which true paths must be identified.
This is usually done in the first instance by using the sequence reads
to disambiguate paths.
In the second instance, this is done by using paired sequence reads
(e.g. \textit{paired-end} and \textit{mate-pair} sequence reads),
in a process usually called \textit{scaffolding}.
The algorithms described in the literature can either be implemented
directly on our representation, or in most cases, adapted.
One important caveat is that our representation depends on the
properties of the de~Bruijn graph (i.e. the relationship between
nodes and the edges that connect them), and while edges may be
added or removed, the representation cannot be treated as an arbitrary
graph -- duplicating or arbitrarily merging parts of the graph.
We do not believe this is a significant obstacle to building a
complete assembler (which we are doing) based on this representation.

As well as building a practical assembler based on the representation
we have presented, there are several opportunities for improving
the graph construction. At the moment, the run-time is dominated by
sorting, which is done sequentially, and with fairly generic sorting
code. We speculate that the sequential sorting speed could be doubled
with modest effort, and the whole could be parallelized fairly easily.

\subsection{A Succinct Representation of Sequence Reads}

Among the several components required for a practical assembler mentioned above,
the use of reads during assembly is worthy of some further examination.
A practical assembler will use the sequence reads to help disambiguate
conflations in the de~Bruijn graph.
Here we present a simple technique that uses succinct data structures
to form a compact representation of the sequence reads, given the
de~Bruijn assembly graph.

The de~Bruijn graph already contains most of the information present
in the sequence reads.
Each sequence read corresponds to a path in the de~Bruijn assembly graph.
The information present in the sequence reads that is not present in the
graph is: (\textit{i}) where in the graph the sequence read starts;
(\textit{ii}) where in the graph it ends, or its length; and (\textit{iii})
at nodes in the graph where there is more than one out-going edge,
which edge should be followed.

If we sort the sequence reads
(discarding the \textit{order} of the reads),
we can efficiently store the initial $k$-mer of each read, and, moreover
construct an efficient index that lets us determine which reads begin with
a given $k$-mer.
The lengths of the reads can be stored efficiently by creating a sparse
bitmap corresponding to the concatenation of all the sequence reads,
with a \textbf{1} denoting the start of a sequence read
(\textbf{rrrarray} would be a logical choice for such a bitmap).
The \textit{rank} and \textit{select} functions give an efficient means
of determining the position in the bitmap of the start and end of a given read.

The sequence of choices, or the \textit{path} that the sequence read follows
may be encoded very efficiently in the following way.
At each node, we can number the extant out-going edges $[0,3]$,
and assign a rank to the edge taken by a given sequence read.
The ranks may be assigned lexicographically, or in order of edge count
(highest to lowest).
These ranks require two bits, which we can store in a pair of sparse
bitmaps -- one for the least significant bit,
and one for the most significant bit.
The positions in these bitmaps correspond to the positions in the bitmap
marking the initial positions of sequence reads.
In practice, a large majority of nodes have only one out-going edge,
so the rank will be~0, hence the bitmaps will be sparse.
Most of the nodes which have more than one out-going edge have only two,
so in the vast majority of cases, the most significant bit of the rank
will be zero, making the bitmap for the most significant bit even more
sparse than the one for the least significant bit.

If one wished to use this encoding to encode sequence reads other than
those represented in the de~Bruijn assembly graph, then it is no longer
the case that every sequence read corresponds to a path in the graph.
In this case, a ``nearest'' path could be found, and the differences
between the sequence read and the path could be recorded. This could
be done using a sparse bitmap to record those positions at which the
path and the sequence read diverge, and a corresponding vector
(indexed by rank in the said bitmap) of bases could be used to store the
actual base in the sequence read.
There is an optimization problem to find the ``nearest'' path,
but simple heuristics are likely to be sufficient.

This scheme could be generalized for sequencing technologies where
we may wish to explicitly encode gaps in the sequence read,
for example \textit{strobe reads} (\cite{strobe}), by the use of an
auxiliary bitmap marking the locations of the gaps.
This would be an interesting line for further research.

\section{Conclusion}

We have presented a memory-efficient representation of the de~Bruijn
assembly graph using \textit{succinct} data structures which allow
us to represent the graph in close to the minimum number of bits.
We have demonstrated its effectiveness on a number of genomes from
bacterial to mammalian scale, including the human genome.
Further work will build on this to produce a practical assembler.

\section*{Acknowledgement} %
Thanks are due to Justin Zobel, Arun S. Konagurthu and Bryan Beresford-Smith
for many fruitful discussions during the long gestation of this
work, and for their feedback on drafts of this paper.

\paragraph{Funding\textcolon} %
National ICT Australia (NICTA) is funded by the Australian Government's
Department of Communications, Information Technology and the Arts,
the Australian Research Council through Backing Australia's Ability,
and the ICT Centre of Excellence programs.

\bibliographystyle{natbib}
\bibliography{document}

\end{document}